\documentclass[aps,prx,reprint,showpacs,floatfix,superscriptaddress]{revtex4-2}
\usepackage{xcolor}
\usepackage{microtype}
\usepackage[bookmarks=true,
bookmarksnumbered=false, 
bookmarksopen=false, 
colorlinks=true,
linkcolor=blue]{hyperref}
\definecolor{webblue}{rgb}{0, 0, 0.5} 
\hypersetup{colorlinks=true,urlcolor=blue,citecolor=blue}
\usepackage{graphicx}
\usepackage{subfigure}
\usepackage{url}
\usepackage{hyperref}
\usepackage{amsmath}
\usepackage[capitalize]{cleveref}
\usepackage{braket}
\usepackage{outlines}
\usepackage{enumitem}
\usepackage{soul}
\usepackage{color}
\usepackage{bm} 

\usepackage[utf8]{inputenc}
\usepackage{multirow}
\usepackage{amssymb}

 \usepackage{relsize}
 
  \usepackage{footnote}

\usepackage{soul}

\usepackage{pgfplots}
\usepackage{ifthen}

\begin{document}




\title{Dynamical screening and excitonic bound states in biased bilayer graphene}

\author{Harley D. Scammell}
\affiliation{School of Physics, the University of New South Wales, Sydney, NSW, 2052, Australia}

\author{Oleg P. Sushkov}
\affiliation{School of Physics, the University of New South Wales, Sydney, NSW, 2052, Australia}

\date{\today}

\begin{abstract}
 Excitonic bound states are characterised by a binding energy $\epsilon_b$ and a single-particle band gap $\Delta_b$. This work provides a theoretical description for both strong  ($\epsilon_b\sim\Delta_b$) and weak ($\epsilon_b\ll\Delta_b$)  excitonic bound states, with particular application to biased bilayer graphene. Standard description of excitons is based on a wave function that is determined by a Schr\"odinger-like equation with screened attractive potential. The wave function approach is valid only in the  weak binding regime $\epsilon_b\ll\Delta_b$. The screening depends on frequency (dynamical screening) and this implies retardation. In the case of strong binding, $\epsilon_b\sim\Delta_b$, a wave function description  is not possible due to the retardation. Instead we appeal to the  Bethe-Salpeter equation, written in terms of the electron-hole Green's function, to solve the problem. So far only the weak binding regime has been achieved experimentally. Our analysis demonstrates that the  strong binding regime is also possible and we specify conditions in which it can be achieved for the prototypical example of biased bilayer graphene. The conditions concern the bias, the configuration of gates, and the substrate material. To verify the accuracy of our analysis we compare with available data for the weak binding regime. We anticipate applying the developed dynamical screening Bethe-Salpeter  techniques to various 2D materials with strong binding.
\end{abstract}

\maketitle


\section{Introduction}

Graphene layers, and the manipulation thereof, are the model hunting ground for peculiar single-particle quantum phases of matter, such as topological insulators, as well as many-body phases, including superconductivity. An important many-body phase actively pursued in graphene layers is the excitonic insulator \cite{KeldyshKopaev1965, RiceKohn1967, Halperin1968} --- a many-body ground state comprising condensed particle-hole pairs. This phase holds promise for novel superfluidity that could be harnessed for low-energy technology \cite{Lozovik1975, pogrebinskii1977mutual, blatt1962bose, kellogg2004vanishing, su2008make}.
Understanding of an isolated exciton is a necessary step for understanding the
exciton condensation.

Exciton is a particle hole bound state in a band  insulator.
Excitons in biased bilayer graphene (BBG) have been observed several years ago
\cite{Ju2017}.
Theoretically the exciton problem in BBG has been considered in Refs. \cite{Park2010,Appelbaum2019,Sauer2022,Henriques2022}. These works ultimately employ the instantaneous screened
Coulomb approximation to find the binding energy and the wave function of the exciton. Often this approach is referred to as the Bethe-Salpeter equation (BSE), however, it is necessary to clarify the terminology: for an instantaneous interaction, a Hamiltonian approach is valid, known generically as the Lippmann-Schwinger equation (LSE). In the case of retardation the approach is the BSE. The distinction is important, the LSE provides a relatively simple wavefunction description, whereas for the BSE a wavefunction is not possible, and instead the correct object is two-particle
Green's function \cite{berestetskii1982quantum}.


In this work we address the issue of retardation in electron-hole binding  in BBG.
There are two main parameters in the problem, exciton binding energy $\epsilon_b$ and  single-particle band gap
$\Delta_b$ induced by bias. 
In the weak binding limit, $\epsilon_b \ll \Delta_b$, retardation is negligible. 
This is the limit addressed in the existing experiment \cite{Ju2017} as well in previous theoretical works \cite{Park2010,Appelbaum2019,Sauer2022,Henriques2022}.
Contrary to this, we find that in the case of strong binding, $\epsilon_b \sim \Delta_b$,
the retardation is non-negligible, and acts to significantly enhance the binding energy.  Notably, the strong binding regime is essential to understand the possibility of exciton condensation. However, we leave the pursuit of condensation for future work.
The importance of retardation in some two dimensional semiconductor exciton problems has been previously pointed out
in Ref. \cite{Glazov2018}. The authors of Ref. \cite{Glazov2018} replace the zero frequency in the
screened potential by some effective frequency, and ultimately solve the LSE. 

Full solution of the BSE is numerically challenging, and presents a bottle-neck. In this work we develop a systematic method to account for retardation, and at a low numerical cost. This is achieved through a perturbative expansion of the BSE. Employing such techniques to the case of BBG, our analysis demonstrates that the strong binding regime in BBG is possible and we specify conditions in which it can  can be achieved. The conditions concern the bias, the configuration of gates, and the substrate material.

    In BBG, the single-particle band gap $\Delta_b$ is proportional to a perpendicular (to the BBG plane) displacement field, which
    is generated via metallic gates above and below the plane, e.g. \cite{Zhang2009}.
    We will assume that the gates are symmetrically placed. Screening, in general, has a significant affect on the
    excitonic binding energy, $\epsilon_b$. There are three sources of screening: (i) dielectric due to a material between BBG and the gates,
    (ii) metallic gates, (iii) in-plane RPA. 
    We will see that to get to the strong coupling regime, $\epsilon_b\sim \Delta_b$, it is necessary to eliminate the dielectric
    material and use suspended BBG -- suspended BBG has been experimentally achieved \cite{yacoby2010,Freitag2012}. Typical energies that we consider are below $100$ meV. Within this range the gate metallic
    screening is practically  frequency independent, but (iii) is frequency dependent. This is the origin of the effect
    that we address.
    
The rest of the paper is structured as follows: In Section \ref{s:singleparticle} we establish the the single particle Hamiltonian and the behaviour of the screened Coulomb interaction. Section \ref{s:Lippmann} introduces the LSE approach, which allows for a particularly straightforward treatment of the two-body exciton problem, without account of retardation (from screening). Section \ref{s:Retard} moves onto the BSE, which accounts for retardation (due to dynamical screening). The approach is more demanding numerically, so we develop a  perturbative expansion which allows for a relatively simple numerical implementation. We find that with account of retardation, binding energies are significantly enhanced and we predict that the strong binding/exciton condensation is possible to implement experimentally. To verify our techniques, in Section \ref{s:Exp} we analyse existing experimental data for weakly bound excitons in BG. Without introducing fitting parameters, we show excellent quantitative agreement. We also resolve a set of unanswered questions. Finally, we discuss our findings and their relation to future experiments in Section \ref{s:disc}.


\section{Single particle Hamiltonian and screened Coulomb interaction.}\label{s:singleparticle}
\subsection{Single particle Hamiltonian}
We will see that the spatial size of the exciton is about $r\sim 1/\sqrt{m\Delta_b}$ where $m\sim 3\times 10^{-2}m_e$ is the effective mass,
hereafter we set $\hbar=1$.
Even at a large band gap $\Delta_b=100$meV the size is about 5nm. Therefore, the continuum approximation is sufficient for
analysis of the problem.
The low-energy single particle Hamiltonian of BBG is \cite{Falko2006}
\begin{eqnarray}
\label{H1}
  H_{0}=\left(
  \begin{array}{c c}
    \Delta & -\frac{\Pi_-^2}{2m}\\
 -\frac{\Pi_+^2}{2m} &-\Delta
  \end{array}
  \right).
\end{eqnarray}
It is written in terms of $\{A_1, B_2\}$ orbitals, with $A,B$ referring to graphene sublattice and subscripts $1,2$ referring to layers. 
Here $\Pi_\pm = \tau (p_x-eA_x) \pm i (p_y - eA_y) \equiv  \tau \Pi_x \pm i \Pi_y$, $\bm p$ is the in-plane momentum, $\bm A$ the
magnetic vector potential,  $\tau=\pm 1$
the valley quantum number, $m$ the effective mass, and $\Delta=\Delta_b/2$, which is proportional to the bias electric field \cite{Zhang2009}.

There are known corrections to this Hamiltonian \cite{McCann2013}, which we gather as a perturbation,
\begin{align}
\label{deltaH}
\delta H&=\begin{pmatrix} 
\frac{\Pi_- \Pi_+}{2M}\left(1-\frac{\Delta}{\Delta_0}\right)&\frac{\Pi_0 \Pi_+}{2m}\\
\frac{\Pi_0 \Pi_-}{2m}&
\frac{\Pi_+ \Pi_-}{2M}\left(1+\frac{\Delta}{\Delta_0}\right).
 \end{pmatrix}
\end{align}
This captures three types of perturbations,  $\frac{\Pi_0 \Pi}{2m}$,
$ \frac{\Pi^2}{2M}$, and $\frac{\Pi^2}{2M}\frac{\Delta}{\Delta_0}$.
The first perturbation controls trigonal warping, the second controls particle-hole
asymmetry and the third does not break symmetries, but is nonetheless treated as a small correction.

The single particle band gap $\Delta_b=2\Delta$ encodes the applied displacement field, $D$. The conversion is observed to be approximately linear, with $e D/(2\Delta)\approx 10.4$ nm$^{-1}$ taken from experiment \cite{Zhang2009}.  Meanwhile, all other parameters of the single particle Hamiltonian can be related to the standard Slonczewski-Weiss-McClure (SWM) parameters of
BG \cite{McCann2013}
\begin{eqnarray}
  \label{mM}
  &&m=\frac{2\gamma_1}{3a^2\gamma_0^2}\approx 0.032m_e, \nonumber\\
  &&M=\frac{m}{2\gamma_4/\gamma_0+\Delta^{\prime}/\gamma_1}\approx \frac{m}{0.146}
  \approx 0.22m_e, \nonumber\\
  &&\Delta_0=\frac{\gamma_1}{2}\frac{m}{M}\approx 28meV,\nonumber\\
  &&\Pi_0=\frac{2\gamma_1\gamma_3}{\sqrt{3}\gamma_0^2a}\approx 0.068nm^{-1}.
    \end{eqnarray}
To obtain the numerical values, we have taken the SWM parameters established by the experiment and analysis of Ref. \cite{Kuzmenko2009},
\begin{gather}
\notag a = 2.46 \AA, \ \gamma_0 = 3.16 eV, \  \gamma_1 = 0.381 eV,\\
 \gamma_3 = 0.38 eV, \
 \gamma_4 = 0.14 eV, \ \Delta' = 0.022 eV.
\end{gather}
Taken together, this uniquely fixes the single particle Hamiltonian --- we therefore have not introduced fitting parameters.

The Hamiltonian (\ref{H1}) [with/without the corrections (\ref{deltaH})] determines the dispersion of the valence, $\epsilon_{\bm p}^{(-)}$,
and conduction, $\epsilon_{\bm p}^{(+)}$, bands, as well as the corresponding wave functions
$\psi^{(\pm)}_{\bm p,\tau}$ (spin index is idle, and so is not included).  We will perform calculations for Hamiltonian (\ref{H1}) both with and without the corrections  (\ref{deltaH}) --- we will see that  perturbations influence results rather weakly for the parameter range of interest. Therefore, for presentation
we omit the correction (\ref{deltaH}) everywhere except of the comparison
with experiment.
Excluding the small corrections (\ref{deltaH}), it is convenient to perform the analysis in rescaled units, whereby energy is measured in units of $\Delta$ and momentum in units
of $p_0=\sqrt{2m\Delta}$,
\begin{eqnarray}
  \label{units}
&&  {\overline \epsilon} = \epsilon/\Delta\nonumber\\
&&  {\overline p} = p/p_0=p/\sqrt{2m\Delta}
\end{eqnarray}
In these units, and in zero magnetic field, the Hamiltonian (\ref{H1}) reads
\begin{eqnarray}
\label{H11}
  H_{0}=\left(
  \begin{array}{c c}
   1 & -{\overline p}_-^2\\
 -{\overline p}_+^2 &-1
  \end{array}
  \right).
\end{eqnarray}
The eigenenergies are $\epsilon_q^{(\pm)}=\pm\sqrt{1+{\overline p}^4}$.

\subsection{Screened Coulomb interaction}
\begin{figure}
	\begin{center}
		\includegraphics[angle = 0, width=0.2\textwidth]{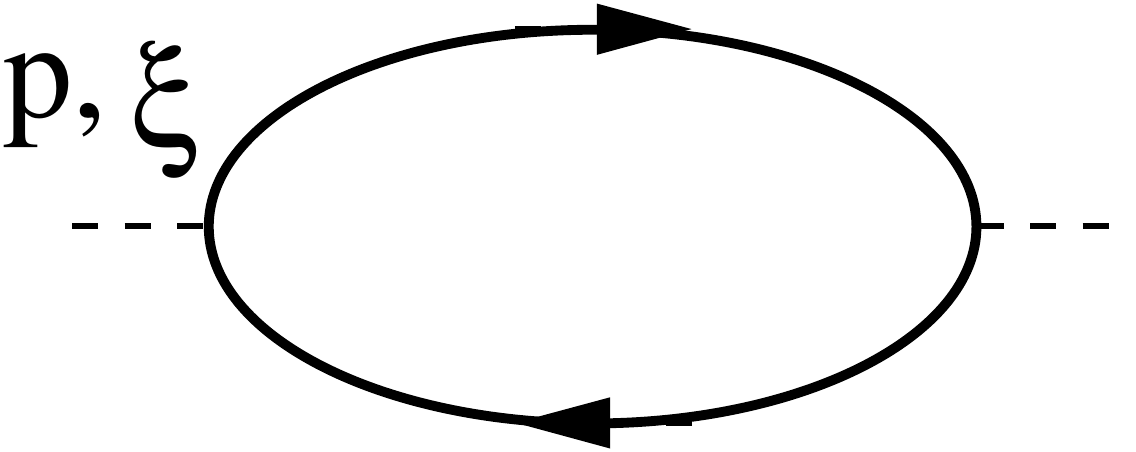}
		\caption{Electron polarisation operator.
}
		\label{PO}
	\end{center}
\end{figure}

\begin{figure}[t!]
  \begin{center}
 	 \includegraphics[angle = 0, width=0.475\textwidth]{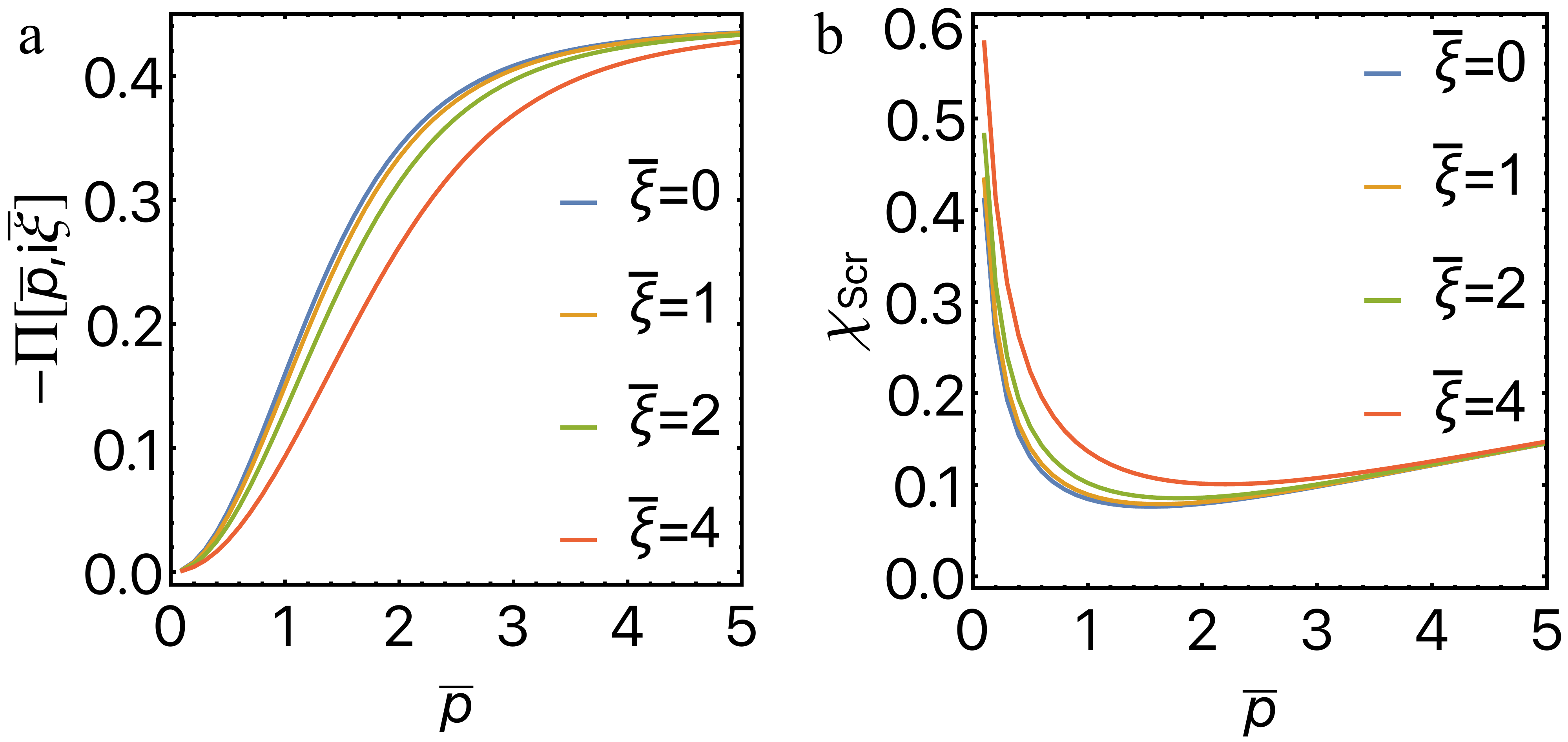}
		\caption{Polarisation operator $-\overline{\Pi}$ (a)
                  and the The Coulomb interaction screening factor
                  $\chi_{\text{scr}}$ (b)  versus ${\overline  p}$ for
                  different values of
                  ${\overline \xi}$.
                  The  screening factor is presented for  $\Delta=15$meV and
                  $\epsilon_r=1$.
                }
		\label{PO1}
	\end{center}
\end{figure}

The many-body RPA  screening of Coulomb interaction in BG is significant. We recall the essential details. The polarisation
operator at an imaginary frequency $\xi$, as given by the diagram in Fig. \ref{PO},
reads
\begin{eqnarray}
  \label{polap}
        \Pi({\bm p},i\xi)=2\times 2\times
        \int D_{\bm q}\frac{2\left(\epsilon_{\bm q}^{(-)}-\epsilon_{\bm {q+p}}^{(+)}\right)F^{+-}_{\bm {q,q+p}}}
    {\left(\epsilon_{\bm q}^{(-)}-\epsilon_{\bm {q+p}}^{(+)}\right)^2+\xi^2}
\end{eqnarray}
Throughout the paper we use the notation
\begin{eqnarray}
D_{\bm q}=\frac{d^2q}{(2\pi)^2}.
\end{eqnarray}
The prefactor $2\times 2$ in (\ref{polap}) comes from spin and valley degeneracy, the vertex factor is the overlap of the conduction and valence single particle wave functions,  
$F^{+-}_{\bm {q,q+p}}=|\langle\psi^{(+)}_{\bm {q+p},\tau}|\psi^{(-)}_{\bm q,\tau}
\rangle|^2$. We do not account for vertices that change the valley; the resulting Coulomb interaction would be significantly suppressed due to the momentum ratio $q/K\ll 1$, where $K$ is the valley momentum. The vertex factor is zero at $\bm q=\bm 0$. The polarisation operator has dimension of mass, rewriting
Eq. (\ref{polap}) in dimensionless units (\ref{units}) gives,
\begin{eqnarray}
  \label{polap1}
\Pi({\bm p},i\xi)=2m \overline{\Pi}({\overline {\bm p}},i{\overline \xi})
\end{eqnarray}
where $\overline{\Pi}$ is given by the same Eq. (\ref{polap}), but all the variables replaced to those with bars. At  ${\overline {q}}\gg 1$ the polarisation operator is
$\overline\Pi=\frac{\ln 4}{\pi}\approx 0.441$, which reduces to the $\Delta=0$ case considered in Ref. \cite{Hwang2008}. In the general case of arbitrary $\Delta$, the operator is straightforwardly computed numerically. 
Plots of $-\overline\Pi$ versus ${\overline  p}$ for different values of
${\overline \xi}$ are presented in Fig. \ref{PO1}a.
The polarisation operator  is zero at ${\overline  p=0}$ and it is approaching
$(\ln 4)/\pi$ at large ${\overline  p}$. 
The frequency dependence of the polarisation operator becomes significant when the 
frequency is comparable and larger than the band gap, $\xi \gtrsim \Delta_b=2\Delta$. 

There are two points to note on the behaviour of the polarisation operator:
 (i) The polarisation screening becomes
significant at ${\overline p} > 1$. The scale ${\overline p} \sim 1$
will determine the size of the exciton, $1/r \sim p_0=\sqrt{2 m \Delta}$; and
(ii) the (imaginary) frequency monotonically reduces screening, i.e.  $\overline{\Pi}({\overline {\bm p}},i{\overline \xi})< \overline{\Pi}({\overline {\bm p}},0)$ for all $\xi>0$. 
We would like to stress that this is true only for imaginary frequency -- for real frequencies the screening properties are complicated and obscured. 
It is convenient to work with imaginary frequency.

Using notation $e^2 = e_0^2/\epsilon_r$, where $e_0$ is the bare charge and $\epsilon_r$ is the dielectric constant, the screened Coulomb interaction is  
\begin{align}
\label{UC}  
V_p(\xi)&=-\frac{2\pi e^2}{p-2\pi e^2\Pi({\bm p},i\xi)}=\frac{1}{2m}{\overline V}_{\overline p}({\overline \xi})\nonumber\\
{\overline V}_{\overline p}({\overline \xi})&=-\frac{4\pi}{ s{\overline p}  
  -4\pi \overline{\Pi}({\overline {\bm p}},i{\overline \xi})}\nonumber\\
s&=\sqrt{\frac{ \epsilon_r^2 \Delta }{m e_0^4/2}}
\end{align}
In these dimensionless units the interaction depends only on the ratio of $\Delta$
over  ``Rydberg'', $Ry=\frac{me^4}{2}$. At $\epsilon_r=1$ the Ry value is
$Ry\approx 435$meV. Hence, for $\Delta=15$meV the parameter $s$ is, $s=0.186$. 
In Fig. \ref{PO1}b we plot the screening factor of the bare Coulomb interaction,
\begin{eqnarray}
  \label{scrf}
  \chi_\text{scr}=\frac{-{\overline V}_{\overline p}}{(4\pi/s{\overline p})}=
  \frac{s {\overline p}}{ s{\overline p}-4\pi \overline{\Pi}({\overline {\bm p}},i{\overline \xi})} \ ,
\end{eqnarray}
  for these parameters. Fig. \ref{PO1}b further emphasises the behaviour seen in Fig. \ref{PO1}a, i.e. that (i) screening is the  most significant at ${\overline p}\sim 1$; and
 (ii) screening is significantly reduced at high imaginary
 frequency ${\overline \xi}\gg1$.

\section{Lippmann-Schwinger equation (LSE)}\label{s:Lippmann}
We define quasi-momentum, $\bm p$, with respect to the valley minimum (K-point) and only consider
a bound states with total quasi-momentum zero, i.e. electron in valley $\tau$, with momentum $\bm p + \tau \bm K$ pairing with a hole in valley $\tau'$, with total momentum $-\bm p - \tau' \bm K$. This means that the total momentum, i.e. as defined with
respect to the $\Gamma$-point, is zero if electron-hole pair in the
same valley (intravalley exciton)
and it is equal to $\pm 2 \bm K$ if they pair in
different valleys (intervalley exciton). Optically, only the intravalley
exciton can be excited.

LSE is a result of summation of ladder diagrams for an instantaneous interaction between the particles. To apply the LSE, we consider the interaction (\ref{UC}), and set $\xi=0$. For the purposes of presentation, we do not include the small corrections \eqref{deltaH}, and therefore $\epsilon^{(+)}_{\bm p}=-\epsilon^{(-)}_{-\bm p}\equiv\epsilon_{\bm p}=\sqrt{\Delta^2+\frac{p^4}{4m^2}}$.
 For some of our numerics, we account for the corrections \eqref{deltaH}. The bound state 
equation reads
\begin{eqnarray}
\label{LS00}
&&(E_0-2\epsilon_{\bm p})\Psi_{\bm p}=\int V_{\bm p-\bm k}(0) Z_{\bm {p,k}}\Psi_{\bm k} D_{\bm k}\\
&&E_0=2\Delta-\epsilon_b^{(0)}\nonumber
\end{eqnarray} 
Here $\Psi_{\bm p}$ is the exciton wavefunction, and $\epsilon_b^{(0)}$ is
the exciton binding energy (the subscript/superscript `0' is used to distinguish from the case with account of retardation, to be discussed in Section \ref{s:Retard}).
The vertex form factors are given by $Z_{\bm {p,k}}^{\tau',\tau}=\langle\psi^{(-)}_{\bm {p},\tau'}|\psi^{(-)}_{\bm k,\tau'}\rangle
\langle\psi^{(+)}_{\bm {k},\tau}|\psi^{(+)}_{\bm p,\tau}\rangle$.  
The form factors do not distinguish spin, yet they weakly distinguish
between intra- and inter-valley excitons.
Note that $\tau$ corresponds to the valley where the electron is located and
$\tau'$ corresponds to the valley where the hole is located.
Explicitly, the expression is
\begin{align}
 Z_{\bm {p,k}}^{\tau',\tau}=
\frac{(1+\Omega_{\bm p}\Omega_{\bm k}e^{2i\tau'\theta})}{1+\Omega_{\bm p}^2}
 \frac{(1+\Omega_{\bm p}\Omega_{\bm k}e^{2i\tau \theta})}{{1+\Omega_{\bm k}^2}}.
\end{align}
Here $\theta=\theta_{\bm k}-\theta_{\bm p}$ and
$\Omega_{\bm p} = 2m(\varepsilon_{\bm p} -\Delta)/p^2$.
Hence, for the intervalley exciton, $\tau'=-\tau$, the Z-factor is
real 
\begin{align}
  \label{Zinter}
  Z_{\bm {p,k}}^{-\tau,\tau}=
\frac{1+\Omega_{\bm p}^2\Omega_{\bm k}^2+2\Omega_{\bm p}\Omega_{\bm k}
  \cos(2\theta)}{(1+\Omega_{\bm p}^2)(1+\Omega_{\bm k}^2)}.
\end{align}
At the same time for the intravalley exciton, $\tau'=\tau$, the Z-factor is
complex
\begin{eqnarray}
  \label{Zintra}
 Z_{\bm {p,k}}^{\tau,\tau}&=&
\frac{1+\Omega_{\bm p}^2\Omega_{\bm k}^2 e^{4i\tau\theta}
  +2\Omega_{\bm p}\Omega_{\bm k} e^{2i\tau\theta}}{(1+\Omega_{\bm p}^2)(1+\Omega_{\bm k}^2)}.
\end{eqnarray}

\begin{figure}[t!]
  \begin{center}
		\includegraphics[angle = 0, width=0.425\textwidth]{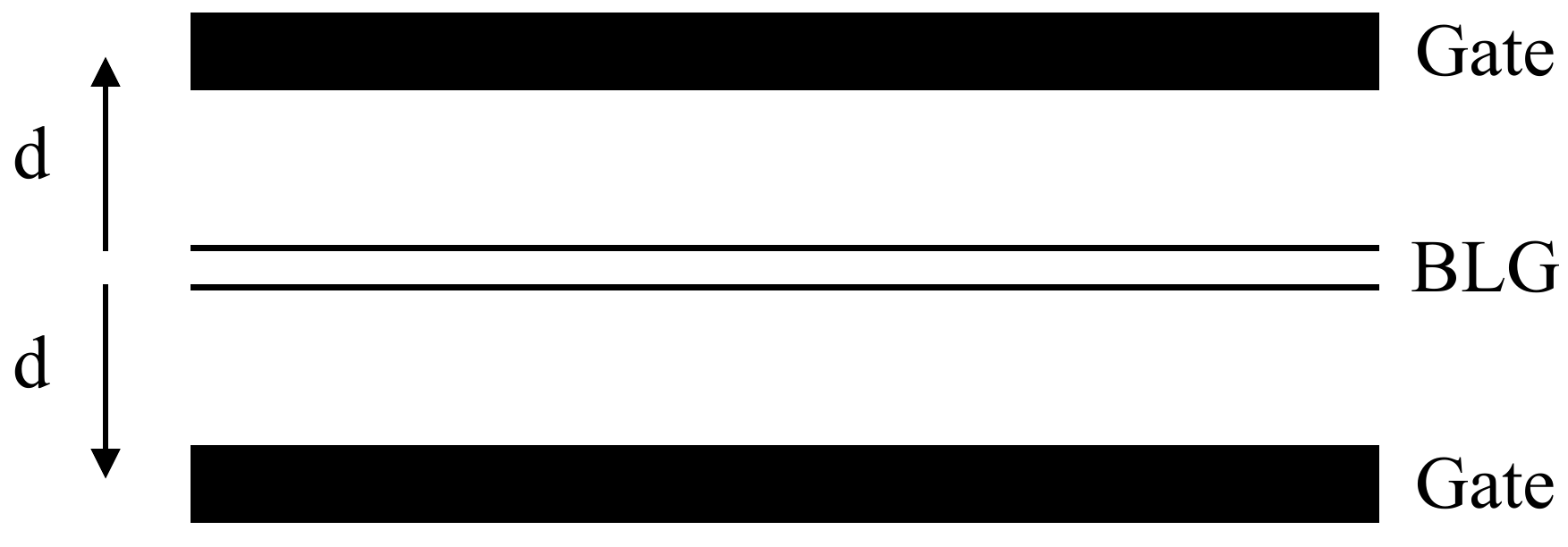}
		\caption{Bilayer graphene with two metallic gates, placed at a distance $d$ above and below.
                }
		\label{setup}
	\end{center}
\end{figure}

In dimensionless units (\ref{units}), Eq.(\ref{LS00}) reads
\begin{eqnarray}
\label{LS01}
({\overline E}_0-2{\overline\epsilon}_{\overline{\bm p}})\Psi_{\overline{\bm p}}&=
\int {\overline V}_{\overline{\bm p}-{\overline{\bm k}}}(0) Z_{\bm {{\overline p},{\overline k}}}\Psi_{\overline{\bm k}}
D_{\overline{\bm k}}.
\end{eqnarray}
The interaction ${\overline V}$ is defined in \eqref{UC}. For brevity, we suppress valley indices on the $Z$-factors. 
As already pointed out, the solution of \eqref{LS01} depends only on the dimensionless parameter $s$, defined in \eqref{UC}.
The interaction in \eqref{UC} has not accounted an important effect --- screening due to the metallic gates.
We consider a setup shown in Fig. \ref{setup}, whereby the top and bottom gates are a distance $d$ from the distance from the BG plane. Accounting for gate-screening, via the method of images, the interaction in Eq.(\ref{UC}) is replaced by,
\begin{eqnarray}
\label{UC1}  
&&{\overline V}_{\overline p}({\overline \xi})=-\frac{4\pi}{ s{\overline p}/\Upsilon_{\bar{p}}
  -4\pi \overline\Pi({\overline {\bm p}},i{\overline \xi})}\nonumber\\
 &&\Upsilon_{\bar{p}}= \tanh(pd)=\tanh(p_0{\overline p}d).
  \end{eqnarray}

Without account of the small trigonal warping $\propto\Pi_0$ in \eqref{deltaH}, we may classify excitonic states $\Psi_{\bm p}$ via 2D angular harmonics
$e^{i\ell\theta_p}$, using
\begin{align}
\label{ell}
\Psi_{\overline {\bm p}} = \sum_\ell 
\frac{1}{\sqrt{\overline p}} \ \psi^\ell_{\overline p} \
e^{i\ell \theta_{\overline p} }\ ,
\end{align}
where $\psi^\ell_{\overline p}$ depends only on the absolute value of momentum.
In a channel with a given orbital momentum $l$ Eq.(\ref{LS01}) is reduced to
\begin{eqnarray}
\label{LS02}
&&({\overline E}_0-2{\overline\epsilon}_{\overline p})\psi^\ell_{\overline{ p}}=
\int_0^{\infty} V^\ell_{{\overline p},{\overline k}} \ \psi^{\ell}_{\overline k} \ d{\overline k}\\
&&V^\ell_{{\overline p},{\overline k}}=
\frac{\sqrt{p k}}{(2\pi)^2}\ \int_0^{2\pi} e^{-i\ell \theta_{\overline p} } \
{\overline V}_{\overline{\bm p}-{\overline{\bm k}}}(0) \ Z_{\bm {{\overline p},{\overline k}}} \
e^{i\ell \theta_{\overline k} } \ d\theta_{\overline k}. \nonumber
\end{eqnarray}
Note, $V^\ell_{{\overline p},{\overline k}}$ is independent of $\theta_{\overline p}$,
since the integrand in the second line of (\ref{LS02}) is a function of
$\theta_{\overline k}-\theta_{\overline p}$.

\begin{figure}[t!]
  \begin{center}
  \includegraphics[width=0.475\textwidth]{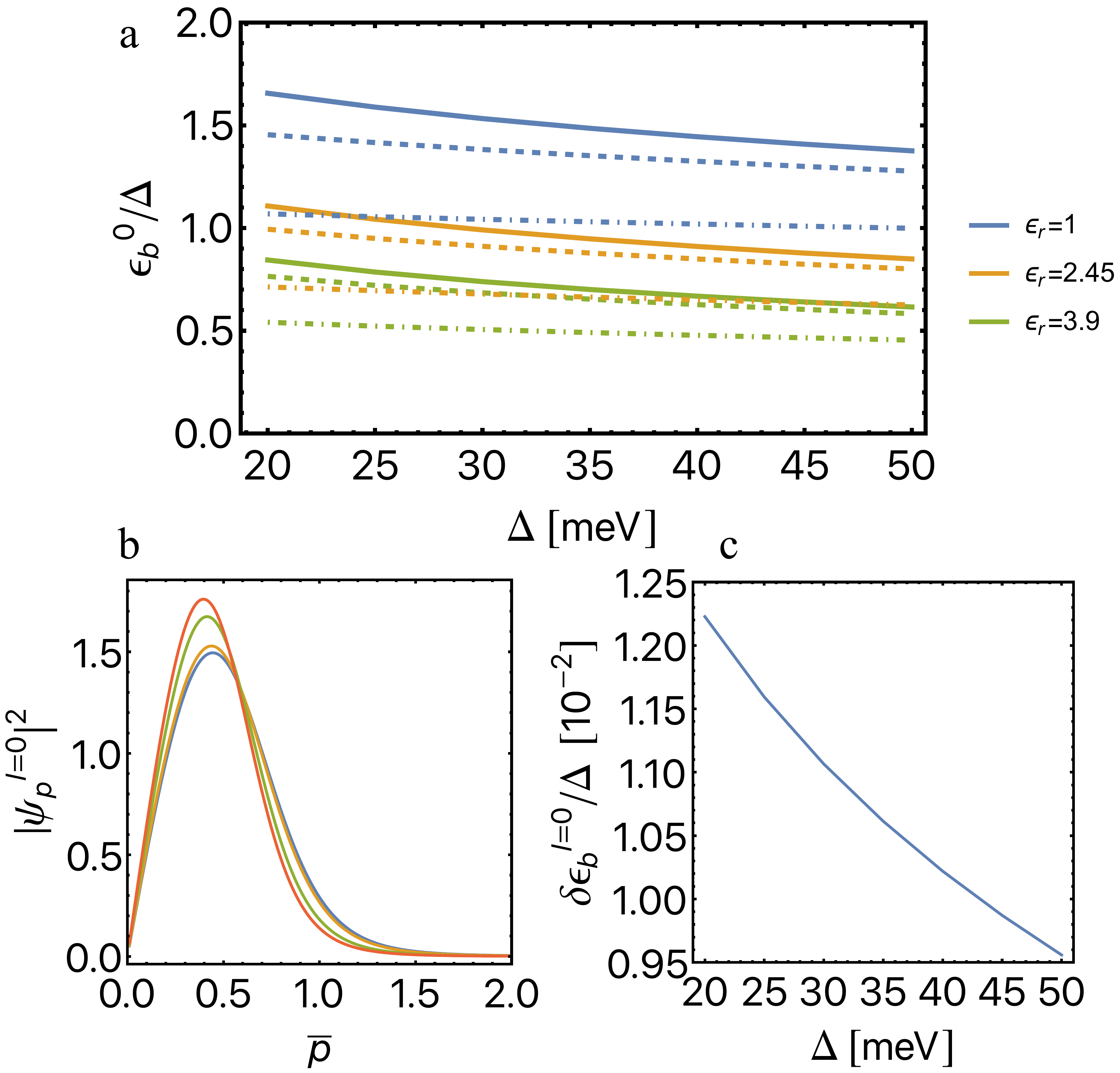}
		\caption{ Exciton s-wave ground state, l=0.
(a)  Binding energies of intervalley exciton in the LSE
approximation versus $\Delta$.
The  plots are presented for three values of the dielectric constant,
$\epsilon_r=1$ (blue curves),
$\epsilon_r=2.45$ (orange curves), $\epsilon_r=3.9$ (green curves), and
three values of the distance to gates $d=1000$nm (solid),
$d=100$nm (dashed), $d=20$nm (dot-dashed).
(b) Wave functions versus dimensionless momentum ${\overline p}$
for different sets of parameters.
The blue line  corresponds to $\Delta=20$meV,
$\epsilon_r=1$, $d=1000$nm; the orange line  corresponds to $\Delta=50$meV, $\epsilon_r=1$, $d=1000$nm;
the green  line  corresponds to $\Delta=20$meV,
$\epsilon_r=3.9$, $d=20$nm;
the red line  corresponds to $\Delta=50$meV,
$\epsilon_r=3.9$, $d=20$nm.
(c) The energy splitting, Eq.(\ref{des}), between the intervalley
and intravalley excitons
versus $\Delta$ for $\epsilon_r=1$ and $d=1000$nm.
	        }\label{bind}
	\end{center}
\end{figure}

Brute force numerical solution of \eqref{LS02} is straightforward.
We consider three cases: (i) suspended BG with the dielectric constant
$\epsilon_r=1$; (ii) single-sided hBN substrate with  effective
$\epsilon_r=(3.9+1)/2=2.45$; and (iii) double-sided hBN substrate with
effective $\epsilon_r=3.9$.
The binding energy of the s-wave ($\ell=0$) ground state of intervalley exciton versus $\Delta$ is
plotted in Fig. \ref{bind}a.
There are nine lines corresponding to the three values of the dielectric constant
and to the three values of the distance to metallic gates, $d=20,100,1000$ nm.
As expected, there is a significant dependence of the binding energy on the dielectric
constant $\epsilon_r$. However, given that the characteristic exciton radius is $r\sim10$ nm, the strong dependence on the gate distance $d>20$nm is somewhat unexpected. Ultimately, this is
because the Coulomb interaction, at zero momentum, is $V_{p\to0}=2\pi e^2 d$. Another surprising observation is practical independence of the wave function on the binding energy. The wave functions corresponding to very different binding
energies are plotted in Fig.\ref{bind}b, and (in dimensionless momenta) are insensitive. 
In the original units they are of course different:
${\overline p}=1$ corresponds to $p=0.130nm^{-1}$ for $\Delta=20$meV and to
$p=0.206nm^{-1}$ for $\Delta=50$meV.  We attribute this universal behaviour to the shape of the polarisation operator, which is small $\propto {\overline p}^2$ up to a scale ${\overline p}=1$.

Due to the difference of the Z-factors, Eqs.(\ref{Zinter}),(\ref{Zintra}), binding energies of intervalley and intravally excitons
are slightly different. The intervalley exciton has a stronger binding.
The difference of binding energies,
\begin{eqnarray}
  \label{des}
\Delta \epsilon_b^s=  \epsilon_b^{(0)}(inter)-\epsilon_b^{(0)}(intra) \ ,
\end{eqnarray}  
is plotted versus $\Delta$ in Fig.\ref{bind}c
for $\epsilon_r=1$ and $d=1000$nm.
The difference in binding energies is less than 1\%. It is even smaller
for higher $\epsilon_r$ and lower d.

\begin{figure}[t!]
  \begin{center}
      \includegraphics[width=0.475\textwidth]{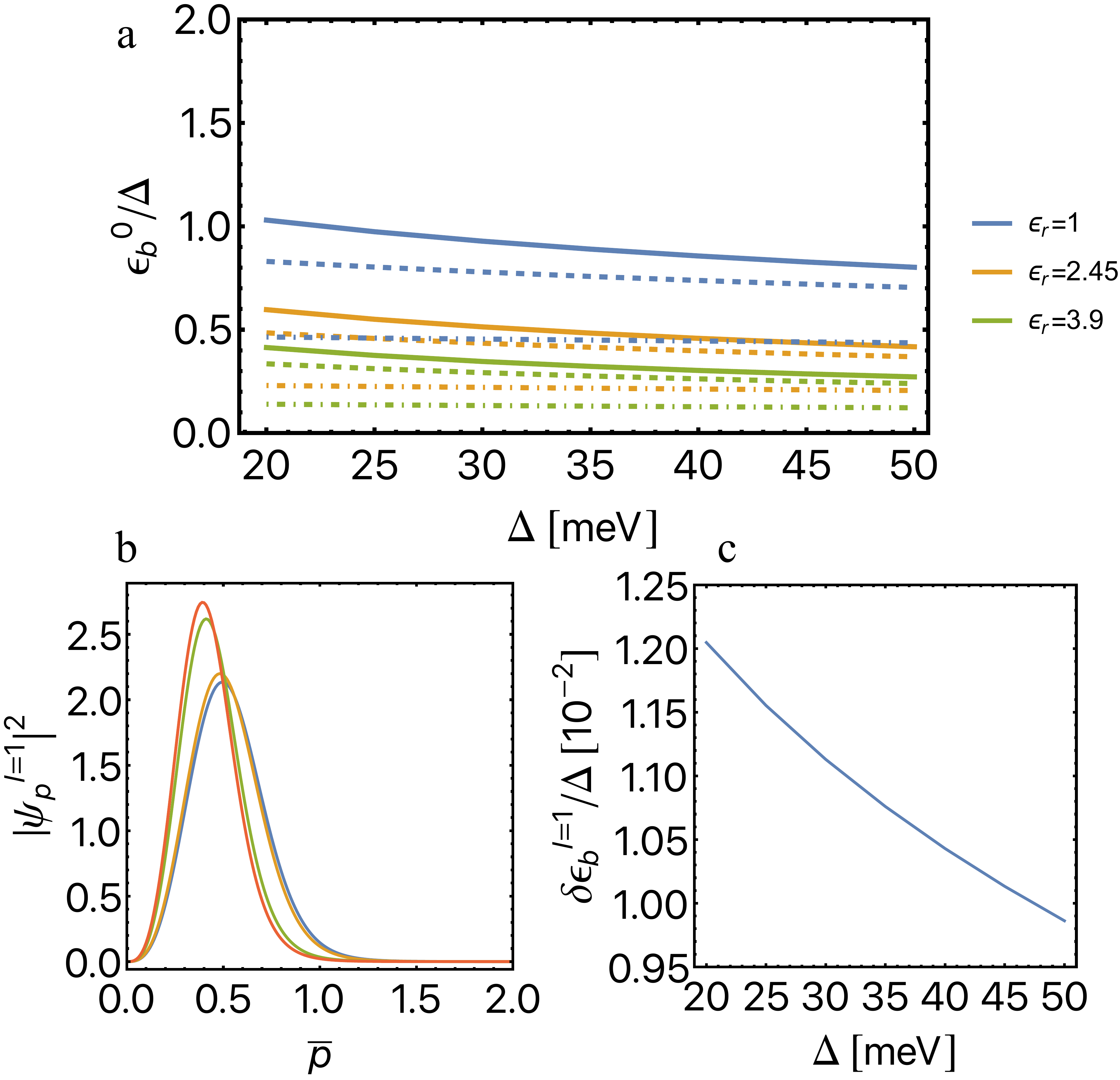}
       \caption{ Exciton lowest $p$-wave  state.
(a) Binding energies of intervalley exciton in the LSE versus $\Delta$.
The  plots are presented for there values of the dielectric constant,
$\epsilon_r=1$ (blue curves),
$\epsilon_r=2.45$ (orange curves), $\epsilon_r=3.9$ (green curves), and
three values of the distance to gates $d=1000$nm (solid),
$d=100$nm (dashed), $d=20$nm (dashed-dotted).
(b) Wave functions versus dimensionless momentum ${\overline p}$
for different sets of parameters.
The blue line corresponds to $\Delta=20$meV,
$\epsilon_r=1$, $d=1000$nm; the orange line  corresponds to $\Delta=50$meV, $\epsilon_r=1$, $d=1000$nm;
the green  line  corresponds to  $\Delta=20$meV,
$\epsilon_r=3.9$, $d=20$nm;
the red line  corresponds to $\Delta=50$meV,
$\epsilon_r=3.9$, $d=20$nm.
(c) The l-dependent, Eq.(\ref{dep})  energy splitting between the
intervalley and intravalley excitons
versus $\Delta$ for $\epsilon_r=1$ and $d=1000$nm.
	        }\label{bindp}
	\end{center}
\end{figure}

For the $p$-wave exciton we again start from the intervalley case.
In this case the energies of $l=\pm 1$ states are degenerate.
The binding energy of lowest  $p$-wave  state of the intervalley exciton is
plotted in Fig.\ref{bindp}a versus $\Delta$ for the same values
of $\epsilon_r$ and $d$ as that for the s-wave in Fig.\ref{bind}.
The wave functions corresponding to different binding
energies are plotted in Fig.\ref{bindp}b.

We already pointed out that the inervalley $p$-wave exciton states
with $l=\pm 1$ are degenerate. 
At the same time the intravalley $p$-wave states with $l=\pm 1$ are nondegenerate, and the following relation is valid
\begin{eqnarray}
  \label{dep}
\epsilon_b^{(0)}(intra)=\epsilon_b^{(0)}(inter)+\tau l  \Delta \epsilon_b^p \ ,
\end{eqnarray}
where $\tau$ indicates the valley.
The splitting $\Delta \epsilon_b^p $ is  plotted in Fig.\ref{bindp}c versus $\Delta$ for $\epsilon_r=1$ and $d=1000$nm.
The splitting is again less than 1\%. It is even smaller
for higher $\epsilon_r$ and lower d.

\section{Retardation and Bethe-Salpeter equation (BSE)}\label{s:Retard}

Now we proceed to the central message of this work.
In the analysis of the previous section, an important effect was neglected
--- retardation of the screened Coulomb interaction.
According to Fig. \ref{PO1}, screening of the Coulomb attraction is monotonically reduced with frequency.
We already pointed out and we would like to stress again that this is true
only for imaginary frequency.
Analytic continuation to real frequency obscures this simple behaviour. It is therefore convenient and physically transparent to
work with imaginary frequency.
Reduced screening enhances
the binding energy.
Heuristically, the typical frequency is set by the binding energy, and therefore when the binding energy is
much smaller than the band gap, $\epsilon_b \ll 2\Delta$, the effect  of screening reduction
is small. However, at strong binding, $\epsilon_b \sim 2\Delta $,
the effect of frequency dependence (retardation) becomes significant.

With account of retardation, the electron and
hole interact at different times and the bound state cannot be described by a wave function. Instead of the wave function, the correct object is the amputated two-particle 
Green's function $\chi_{\xi,{\bm p}}$ -- written here in terms of the relative momentum ${\bm p}$
and the relative frequency $\xi$. In our analysis, the total momentum of the electron and hole is encoded in the valley indices, and 
is either zero  (intravalley pairing) or $2\bm K$ (intervalley pairing).
The imaginary frequency $\xi$ is Fourier conjugated to the retardation time (plus a Wick rotation).
BSE for $\chi_{\xi,{\bm p}}$ reads \cite{berestetskii1982quantum},
\begin{align}
\label{BSe}
\notag \chi_{\xi,\bm p}&=-\frac{1}{(E/2-\omega_{\bm p})^2+\xi^2}\int V_{\bm p, \bm k}(\xi-\lambda)\chi_{\lambda,\bm k}  Z_{\bm {p,k}} D_\lambda D_{\bm k},\\
D_\lambda&=\frac{d\lambda}{2\pi}, \ D_{\bm k}=\frac{d^2k}{(2\pi)^2}. 
\end{align}
Here $E=2\Delta -\epsilon_b$, and $\xi, \lambda$ are imaginary frequencies. 
If the interaction is independent of frequency, $V_{\bm p, \bm k}(\xi-\lambda) \to V_{\bm p, \bm k}(0)$,
BSE \eqref{BSe} is equivalent to LSE \eqref{LS00}, and $\chi$ is related to the usual wave function $\Psi_{\bf p}$ as
\begin{eqnarray}
  \label{chi0}
&&  \chi^{(0)}_{\xi,{\bf p}}= \frac{2a_{\bf p}}{a_{\bf p}^2+\xi^2}
  \Psi_{\bf p},\nonumber\\
&&a_{\bm p}=-E_0/2+\omega_{\bf p}.
  \end{eqnarray}
The superscript/subscript  ``0"  indicates that this is the solution without retardation.
Note that $a_{\bm p}$ is always positive.

\subsection{Perturbation theory for the retardation effect}
Eq.(\ref{BSe}) is not a linear eigenvalue problem, and a direct numerical solution
of Eq.(\ref{BSe}) is an involved  calculation.
Here we develop a perturbation theory method that is sufficient for our purposes.
This is the regime when  the retardation correction while being important is still relatively
small. Let us first replace the interaction in (\ref{BSe}) by a frequency independent interaction,
$V_{\bm p, \bm k}(\xi) \to V_{\bm p, \bm k}^{(0)}$. It can be the interaction at zero frequency,
$V_{\bm p, \bm k}^{(0)}=V_{\bm p, \bm k}(\xi=0)$, or interaction at some typical frequency
$V_{\bm p, \bm k}^{(0)}=V_{\bm p, \bm k}(\xi=\xi_{typical})$, or something else. We will discuss specific
possibilities later.
BSE with $V_{\bm p, \bm k}^{(0)}$ is reduced to LSE which is a linear eigenvalue
problem and can solved numerically with ease. The solution is given by Eq.(\ref{chi0}) where
$\Psi_{\bf p}$ and $E_0$ is the eigenfunction and the eigenenergy of LSE.
Next, let us consider
\begin{eqnarray}
  \label{dV}
  \delta V_{\bm p,\bm k}(\xi)=V_{\bm p, \bm k}(\xi)-V_{\bm p, \bm k}^{(0)} 
\end{eqnarray}
as a perturbation. We obtain the following expression for the first order retardation correction to the binding energy,
\begin{eqnarray}
  \label{de}
  \delta E=\int (\chi^{(0)}_{\xi,{\bf p}})^{*} \delta V_{{\bm p, \bm k}}(\xi-\lambda)
  \chi_{\lambda,{\bf k}}^{(0)}Z_{\bm {p,k}}  D_{\xi}D_{\bf p}D_{\lambda}D_{\bf k}.
\end{eqnarray}

To derive (\ref{de}) let us represent the Green's function as $\chi_{\xi,{\bf p}}=\chi^{(0)}_{\xi,{\bf p}}
+\delta \chi_{\xi,{\bf p}}$. Hence, BSE (\ref{BSe}) can be rewritten as
\begin{eqnarray}
\label{BSe1}
&& \chi^{(0)}_{\xi,\bm p}+\delta \chi_{\xi,{\bf p}}
=-\frac{1}{(E/2-\omega_{\bm p})^2+\xi^2}\\
&&\times
\int [V_{\bm p, \bm k}^{(0)}+\delta V_{\bm p, \bm k}(\xi-\lambda)]
     [\chi^{(0)}_{\lambda,{\bf k}} +\delta \chi_{\lambda,{\bf k}}]  Z_{\bm {p,k}} D_\lambda D_{\bm k} \nonumber\\
&&\approx-\frac{1}{(E/2-\omega_{\bm p})^2+\xi^2}     \nonumber\\
     &&\times \int [V_{\bm p, \bm k}^{(0)}\chi^{(0)}_{\lambda,{\bf k}}
       +\delta V_{\bm p, \bm k}(\xi-\lambda)\chi^{(0)}_{\lambda,{\bf k}}+
       V_{\bm p, \bm k}^{(0)}\delta \chi_{\lambda,{\bf k}}]  Z_{\bm {p,k}} D_\lambda D_{\bm k}.\nonumber
     \end{eqnarray}
The second order term $\delta V \delta\chi$ has been neglected in the last line.
Let us denote
\begin{eqnarray}
\delta\Psi_{\bm p}=\int  \delta \chi_{\xi,{\bf p}} D_\xi.
\end{eqnarray}
Hence, integrating (\ref{BSe1}) over $\xi$ we get
\begin{eqnarray}
  \label{BSe2}
&&\Psi_{\bm p}+\delta\Psi_{\bm p}=+\frac{1}{E-2\omega_{\bm p}}\int V_{\bm p, \bm k}^{(0)}
Z_{\bm {p,k}}\Psi_{\bm k}D_{\bm k}\nonumber\\
&&-\int \frac{D_{\xi}}{(E/2-\omega_{\bm p})^2+\xi^2}\delta V_{\bm p, \bm k}(\xi-\lambda)
\chi^{(0)}_{\lambda,{\bf k}}
Z_{\bm {p,k}} D_\lambda D_{\bm k}\nonumber\\
&&+\frac{1}{E-2\omega_{\bm p}}\int V_{\bm p, \bm k}^{(0)}Z_{\bm {p,k}}\delta\Psi_{\bm k} D_{\bm k}.\nonumber
\end{eqnarray}
This is equivalent to
\begin{eqnarray}
  \label{BSe3}
&&(E-2\omega_{\bm p})(\Psi_{\bm p}+\delta\Psi_{\bm p})=\int V_{\bm p, \bm k}^{(0)}
Z_{\bm {p,k}}\Psi_{\bm k}D_{\bm k}\nonumber\\
&&+\int\frac{2(-E/2+\omega_{\bm p})}{(E/2-\omega_{\bm p})^2+\xi^2}\delta V_{\bm p, \bm k}
(\xi-\lambda)\chi^{(0)}_{\lambda,{\bf k}}
Z_{\bm {p,k}} D_{\xi} D_\lambda D_{\bm k}\nonumber\\
&&+\int V_{\bm p, \bm k}^{(0)}Z_{\bm {p,k}}\delta\Psi_{\bm k} D_{\bm k}.\nonumber
\end{eqnarray}
Representing $E=E_0+\delta E$ and neglecting all the second order terms,
$O(\delta \times \delta)$, this is transformed to
\begin{eqnarray}
  \label{BSe4}
&&\delta E \Psi_{\bm p}+(E_0-2\omega_{\bm p})\delta\Psi_{\bm p}\nonumber\\
&&=\int\frac{2(-E_0/2+\omega_{\bm p})}{(E_0/2-\omega_{\bm p})^2+\xi^2}\delta V_{\bm p, \bm k}
(\xi-\lambda)\chi^{(0)}_{\lambda,{\bf k}}
Z_{\bm {p,k}} D_{\xi} D_\lambda D_{\bm k}\nonumber\\
&&+\int V_{\bm p, \bm k}^{(0)}Z_{\bm {p,k}}\delta\Psi_{\bm k} D_{\bm k}.\nonumber
\end{eqnarray}
Finally, multiplying this Eq. by $\Psi_{\bm p}^*$ and integrating by ${\bm p}$ we
arrive to Eq.(\ref{de}).
Using (\ref{chi0}), one frequency integration in (\ref{de}) can be performed analytically and the retardation correction to the
energy reduces to
\begin{align}
  \label{de1}
\delta E=\int\Psi^{*}_{\bf p} 
 \frac{4(a_{\bm p}+a_{\bm k}) \delta V_{\bm p, \bm k}(\mu)}{(a_{\bm p}+a_{\bm k})^2+\mu^2}\Psi_{\bf k}Z_{\bm {p,k}}
D_{\mu}D_{\bf p}D_{\bf k}.
\end{align}
Here the $\mu$-integration goes from 0 to $\infty$.
Since $\delta V$ is negative the retardation correction to the total energy
is negative, $\delta E<0$, and thereby increases the binding energy. 
If $\delta V$ is independent of frequency, $\delta V_{\bm p,\bm k}(\mu)\to \delta V_{\bm p,\bm k}$  the $\mu$-integration
in (\ref{de1}) is trivial and the energy correction is reduced to the familiar expression from quantum mechanics, $ \delta E=  \int \Psi^*_{\bf p}\delta V_{\bm p, \bm k}\Psi_{\bf k} Z_{\bm {p,k}}D_{\bf p}D_{\bf k}$.

Momenta integrations in (\ref{de1}) are well convergent.
  However, the perturbation $\delta V(\mu)$ increases with frequency, and we
  find that at  large $\mu$ the integrand in (\ref{de1}) decays as $1/\mu$
    leading to a logarithmic divergence of the integral.
To remedy, we impose 
an ultraviolet cutoff $\Lambda=0.5$eV, which corresponds to the energy scale that the single particle Hamiltonian
(\ref{H1}) becomes invalid; one must account for additional bands (i.e. consider the basis $\{A_1,A_2,B_1,B_2\}$) see e.g. \cite{McCann2013}. A change of the cutoff to, say, 1 eV does not significantly influence the retardation correction.

\subsection{Zero frequency RPA potential as starting approximation}
Let us consider first the solution of LSE (\ref{LS00})
obtained in Section \ref{s:Lippmann} as the zeroth approximation.
So, we set $V_{\bm p, \bm k}^{(0)}=V_{\bm p, \bm k}(\xi=0)$.
\begin{figure}[t!]
  \begin{center}
  \includegraphics[width=0.475\textwidth]{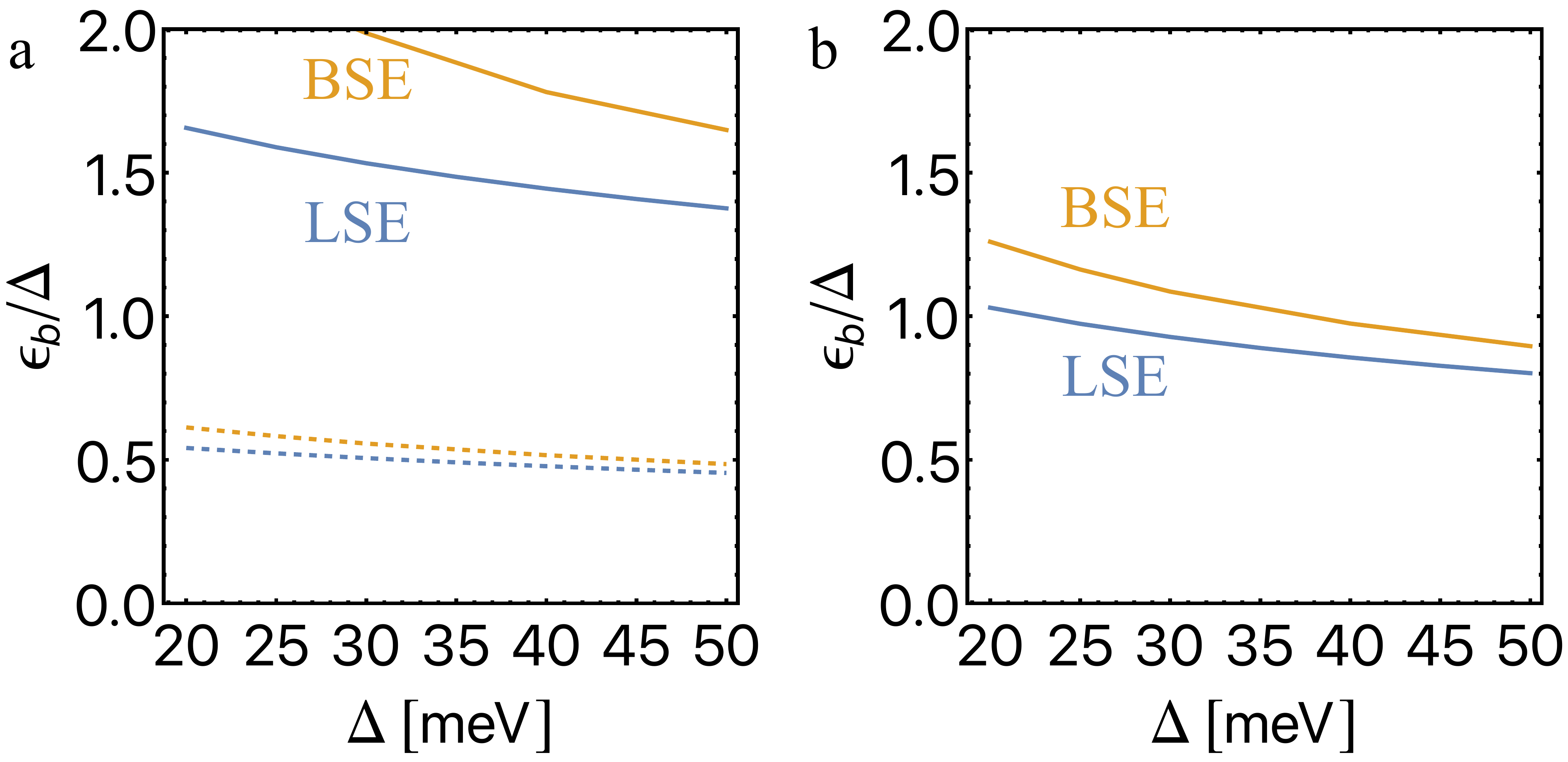}
    \caption{ Exciton binding energies.
      (a) s-wave ground state for two sets of parameters,
      $d=1000$nm, $\epsilon_r=1$ (two upper solid lines)
      and  $d=20$nm, $\epsilon_r=3.9$ (two lower dashed lines).
         (b) Lowest $p$-wave state for $d=1000$nm, $\epsilon_r=1$.
         In both panels LSE (blue curves) is solved with zero
         frequency RPA screened potential. The BSE solution
         includes the retardation correction (\ref{de1}).
       }\label{binds0}
	\end{center}
\end{figure}
In Fig.\ref{binds0}a we present binding energies of the
s-wave ground state for two sets of parameters: (i) $d=1000$nm, $\epsilon_r=1$;
and (ii) $d=20$nm, $\epsilon_r=3.9$. The blue solid and dashed-dotted
lines show the LSE solutions. These lines are identical to that
in Fig.\ref{bind}a. Orange lines, solid and dashed-dotted, show
the same energies with account of the retardation correction (\ref{de1}).
The retardation correction is significant, especially for the
$d=1000$nm, $\epsilon_r=1$. Moreover, it is qualitatively significant
because for sufficiently small $\Delta$ it brings the system to the
exciton condensation regime, $\epsilon_b > 2\Delta$. The exciton condensation
regime will be considered in a separate publication.

In Fig.\ref{binds0}b we present the binding energy of the
lowest $p$-wave state for $d=1000$nm, $\epsilon_r=1$.
The blue solid line shows the LSE solution. This line is
identical to that in Fig.\ref{bindp}a.
The orange line show  the same energy with account of the retardation correction
(\ref{de1}). For the $p$-wave the retardation is less important.
This is natural, as we already pointed out the retardation is more important
for a larger binding energy.

\subsection{Averaged over frequency RPA potential as starting approximation}
\begin{figure}[t!]
  \begin{center}
      \includegraphics[width=0.475\textwidth]{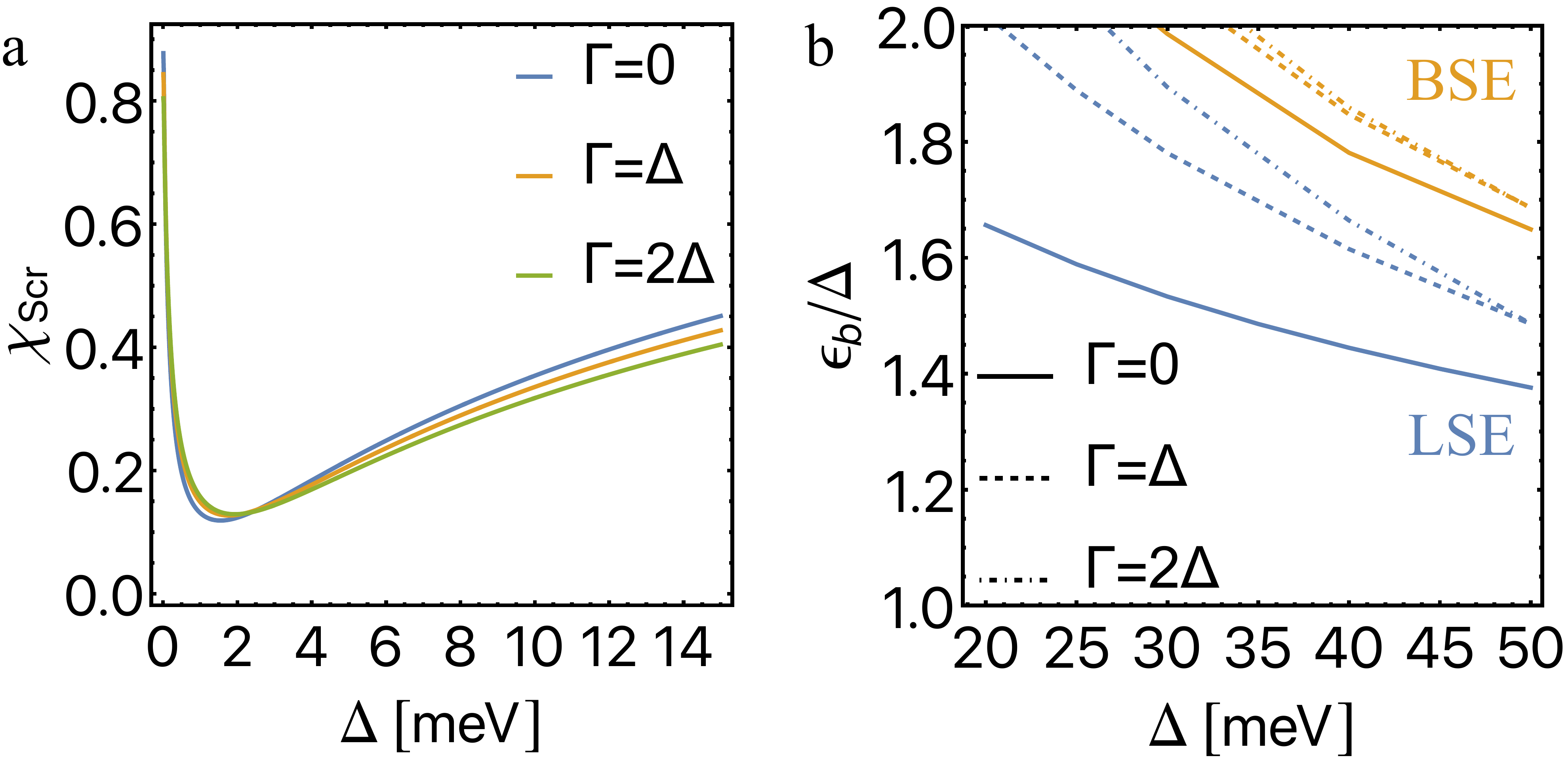}
    \caption{(a) Screening factors defined by Eq.(\ref{scrf})
      for frequency averaged potentials, Eq.(\ref{Uav}),  with
      $\Gamma=0,\Delta,2\Delta$ at  $\Delta=40$meV.
(b)  Binding energies calculated
using the averaged potential method. Blue lines present LSE
binding energies calculated with averaged potentials with $\Gamma=0$
(solid), $\Gamma=\Delta$ (dashed), and $\Gamma=2\Delta$ (dashed-dotted).
The orange lines are the BSE binding energies (LSE
with added retardation correction (\ref{de1})).
}\label{pota}
  \end{center}
\end{figure}

We can improve accuracy of the calculation of the retardation effect.
To do so let us change the zero approximation potential that enters LSE. 
Instead of the zero frequency, $V_{\bm p, \bm k}^{(0)}=V_{\bm p, \bm k}(\xi=0)$,
we take the frequency averaged potential
\begin{eqnarray}
  \label{Uav}
  V_{\bm p, \bm k}^{(0)}=
  \frac{2}{\pi}\int_0^{\Lambda}d\xi \frac{\Gamma}{\xi^2+\Gamma^2}
  V_{{\bm p},{\bm k}}(\xi)
  \end{eqnarray}
Where the frequency dependent potential is given by Eq.(\ref{UC}).
Here $\Lambda=0.5$eV is the ultraviolet cutoff.
The case $\Gamma=0$ corresponds to the static screening considered in the
previous subsection. From discussion in previous sections we suggest
that the optimal value of $\Gamma$ is $\Gamma \approx \epsilon_b$.
It is worth noting that the averaging (\ref{Uav}) makes sense only in
imaginary frequency. This is because the dependence of $V_{{\bm p},{\bm k}}(\xi)$
on $\xi$ is monotonic. A similar averaging in real frequency would have limited meaning.
In Fig.\ref{pota}a
we present plots of the screening factors, Eq.(\ref{scrf}), for averaged
potentials with $\Gamma=0,\Delta,2\Delta$ at  $\Delta=40$meV.
The averaging captures some of the physics or retardation, and thereby reduces screening in the zeroth approximation (i.e. in the LSE approach).

In Fig.\ref{pota}b we present binding energies calculated
using the averaged potential method. Blue lines present LSE
binding energies calculated with averaged potentials with $\Gamma=0$
(solid), $\Gamma=\Delta$ (dashed), and $\Gamma=2\Delta$ (dashed-dotted).
The orange lines are the LSE binding energies with added
retardation correction (\ref{de1}).
Solid lines, black and red, are identical to that in
Fig.\ref{binds0}a.

From results of this subsection we conclude that the appropriate averaging
of the RPA potential over imaginary frequency with subsequent usage
of instantaneous LSE can account up to 50\% of the
retardation correction. For the accurate result one should combine
the potential averaging, Eq.(\ref{Uav}),  and the explicit retardation
correction (\ref{de1}).
We reiterate again, this analysis is important for large binding energy,
$\Delta \lesssim \epsilon_b \lesssim 2\Delta$. For weak binding the
retardation correction is small and the method of the correction calculation
is not very important.

\section{Comparison with existing data}\label{s:Exp}

\begin{figure}[t]
{\includegraphics[width=0.4\textwidth,clip]{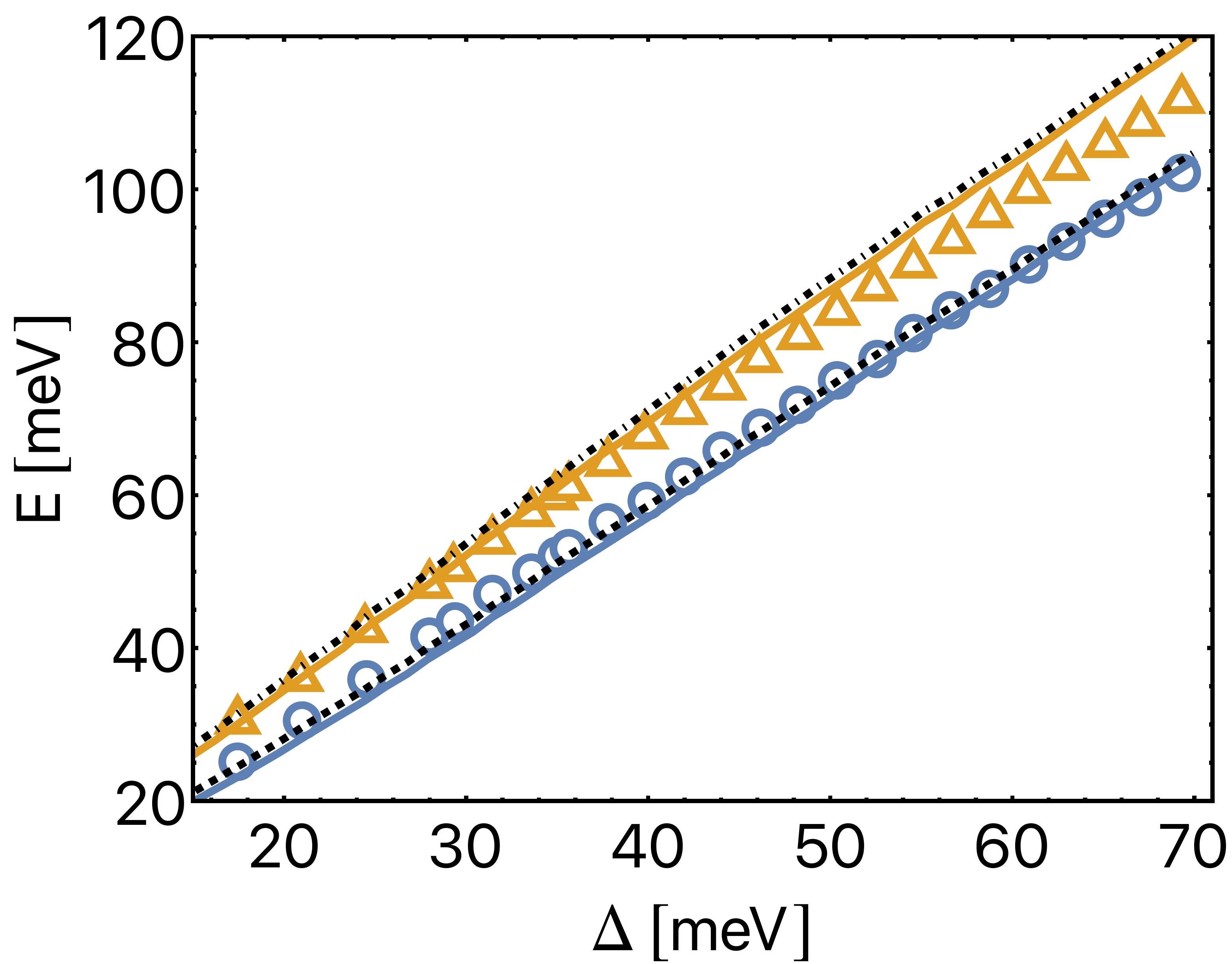}}
\caption{Exciton energies for $s$- and $p$-wave states. Experiment for $s$-wave ($p$-wave) is shown by blue circles (orange triangles). LSE solutions [including $\delta H$ \eqref{deltaH}]  for $s$- and $p$-wave are given by dashed and dot-dashed curves. BSE solutions [including $\delta H$ \eqref{deltaH}] is shown by solid blue/orange curves for $s$/$p$-wave. }
\label{f:E_exp}
\end{figure}

Assuming hBN encapsulation, giving dielectric enhancement $\bar{\epsilon}=3.9$, and taking the metallic gates to be at a distance $d=20$nm, we can directly compare with the experimental measurements of Ref. \cite{Ju2017}. We compare the exciton energies, $2\Delta-\epsilon_b$, vs $\Delta$ obtained using LSE and BSE, to those measured experimentally. The experiment measures both $s$- and $p$-wave intravalley excitons, so it is necessary to use form factor \eqref{Zintra}. The comparison is provided in Fig. \ref{f:E_exp}, from which we see that our techniques provide quantitative agreement. We stress that we have not introduced fitting parameters. We reiterate that this is the case of weak binding.

\section{Discussion}\label{s:disc}

{\bf Summary.} 
In this work we considered the influence of dynamical screening (retardation) on exciton binding -- i.e the binding energy $\epsilon_b$. We considered the particular example of biased bilayer graphene, whereby the bias field induces single-particle band gap $\Delta_b$. However, the techniques developed are applicable to many other 2D materials. 

We found that for  $\epsilon_b \ll \Delta_b$, retardation effects can be safely neglected, and the properties of the exciton bound states are very well captured by the Lippmann-Schwinger equation (a two-body Hamiltonian approach). However, in the strong binding regime $\epsilon_b \sim \Delta_b$ retardation is significant and therefore a Hamiltonian approach is insufficient. Instead the correct formalism is the Bethe-Salpeter equation (BSE).  The BSE is costly to numerically implement and thereby presents a  bottleneck. To handle this situation, we develop a simple perturbative expansion of the BSE, which allows us to systematically compute corrections to the binding energy relative to the static case. 

Screening significantly influences the excitonic binding energy. We argue that to probe critical regime of $\epsilon_b \sim \Delta_b$, one must reduce screening from the environment; both dielectric and gate. We propose suspended bilayer graphene (i.e. dielectric $\epsilon_r=1$) with placement of metallic gates $d>20$nm above the plane. Counterintuitively, even though the characteristic radius of excitons considered here is $r\lesssim10$nm, the difference in binding energies for metallic gates at $d=20$ and $100$nm is significant, Fig. \ref{bind}. 

We verify the quantitative accuracy of the methods via directly fitting to available experimental data in this regime \cite{Ju2017}.  Crucially, we take parameters established elsewhere, and as such do not use any fitting parameters.

{\bf Outlook.} 
The crucial finding is the role of dynamical screening, which becomes significant in the regime $\epsilon_b \sim \Delta_b$. To probe this regime, we needed to consider the case of suspended BG with well separated metallic gates to reduce environment screening and thereby maximise the Coulomb interaction. Additionally, one could consider engineering of BG so as to generate a larger effective mass $m$ (smaller bandwidth). All things equal, enhancing the effective mass reduces the kinetic energy and therefore helps to promote condensation. This situation could be achieved via e.g. modulated electrostatic gating or via relative twist of the layers, or other van der Waals engineering. Application of the present techniques to modified BG is an important line of inquiry left for future work.

Finally, we leave it for future work to probe the possibility of exciton condensation, $\epsilon_b\geq\Delta_b$, and characterise the subsequent condensate. 

\section{Acknowledgements}
This work benefited from a range of fruitful discussions with Alex Hamilton, Oleh Klochan, Dmitry Efimkin and Mike Zhitomirsky. We acknowledge funding support from the Australian Research Council Centre of Excellence in Future Low-Energy Electronics Technology (FLEET) (CE170100039).

\bibliography{BLG.bib}

\begin{thebibliography}{22}%
\makeatletter
\providecommand \@ifxundefined [1]{%
 \@ifx{#1\undefined}
}%
\providecommand \@ifnum [1]{%
 \ifnum #1\expandafter \@firstoftwo
 \else \expandafter \@secondoftwo
 \fi
}%
\providecommand \@ifx [1]{%
 \ifx #1\expandafter \@firstoftwo
 \else \expandafter \@secondoftwo
 \fi
}%
\providecommand \natexlab [1]{#1}%
\providecommand \enquote  [1]{``#1''}%
\providecommand \bibnamefont  [1]{#1}%
\providecommand \bibfnamefont [1]{#1}%
\providecommand \citenamefont [1]{#1}%
\providecommand \href@noop [0]{\@secondoftwo}%
\providecommand \href [0]{\begingroup \@sanitize@url \@href}%
\providecommand \@href[1]{\@@startlink{#1}\@@href}%
\providecommand \@@href[1]{\endgroup#1\@@endlink}%
\providecommand \@sanitize@url [0]{\catcode `\\12\catcode `\$12\catcode
  `\&12\catcode `\#12\catcode `\^12\catcode `\_12\catcode `\%12\relax}%
\providecommand \@@startlink[1]{}%
\providecommand \@@endlink[0]{}%
\providecommand \url  [0]{\begingroup\@sanitize@url \@url }%
\providecommand \@url [1]{\endgroup\@href {#1}{\urlprefix }}%
\providecommand \urlprefix  [0]{URL }%
\providecommand \Eprint [0]{\href }%
\providecommand \doibase [0]{https://doi.org/}%
\providecommand \selectlanguage [0]{\@gobble}%
\providecommand \bibinfo  [0]{\@secondoftwo}%
\providecommand \bibfield  [0]{\@secondoftwo}%
\providecommand \translation [1]{[#1]}%
\providecommand \BibitemOpen [0]{}%
\providecommand \bibitemStop [0]{}%
\providecommand \bibitemNoStop [0]{.\EOS\space}%
\providecommand \EOS [0]{\spacefactor3000\relax}%
\providecommand \BibitemShut  [1]{\csname bibitem#1\endcsname}%
\let\auto@bib@innerbib\@empty
\bibitem [{\citenamefont {Keldysh}\ and\ \citenamefont
  {Kopaev}(1964)}]{KeldyshKopaev1965}%
  \BibitemOpen
  \bibfield  {author} {\bibinfo {author} {\bibfnamefont {L.~V.}\ \bibnamefont
  {Keldysh}}\ and\ \bibinfo {author} {\bibfnamefont {Y.~V.}\ \bibnamefont
  {Kopaev}},\ }\href@noop {} {\bibinfo {title} {Possible instability of the
  semimetallic state against coulomb interaction}},\ \bibinfo {howpublished}
  {Fiz. Tverd. Tela., 6, 2791 (1964) [Sov. Phys. Solid State 6, 2219 (1965)]}
  (\bibinfo {year} {1964})\BibitemShut {NoStop}%
\bibitem [{\citenamefont {J\'erome}\ \emph {et~al.}(1967)\citenamefont
  {J\'erome}, \citenamefont {Rice},\ and\ \citenamefont {Kohn}}]{RiceKohn1967}%
  \BibitemOpen
  \bibfield  {author} {\bibinfo {author} {\bibfnamefont {D.}~\bibnamefont
  {J\'erome}}, \bibinfo {author} {\bibfnamefont {T.~M.}\ \bibnamefont {Rice}},\
  and\ \bibinfo {author} {\bibfnamefont {W.}~\bibnamefont {Kohn}},\ }\bibfield
  {title} {\bibinfo {title} {Excitonic insulator},\ }\href
  {https://doi.org/10.1103/PhysRev.158.462} {\bibfield  {journal} {\bibinfo
  {journal} {Phys. Rev.}\ }\textbf {\bibinfo {volume} {158}},\ \bibinfo {pages}
  {462} (\bibinfo {year} {1967})}\BibitemShut {NoStop}%
\bibitem [{\citenamefont {Halperin}\ and\ \citenamefont
  {Rice}(1968)}]{Halperin1968}%
  \BibitemOpen
  \bibfield  {author} {\bibinfo {author} {\bibfnamefont {B.}~\bibnamefont
  {Halperin}}\ and\ \bibinfo {author} {\bibfnamefont {T.}~\bibnamefont
  {Rice}},\ }\bibfield  {title} {\bibinfo {title} {The excitonic state at the
  semiconductor-semimetal transition}\ }(\bibinfo  {publisher} {Academic
  Press},\ \bibinfo {year} {1968})\ pp.\ \bibinfo {pages}
  {115--192}\BibitemShut {NoStop}%
\bibitem [{\citenamefont {Lozovik}\ and\ \citenamefont
  {Yudson}()}]{Lozovik1975}%
  \BibitemOpen
  \bibfield  {author} {\bibinfo {author} {\bibfnamefont {Y.~E.}\ \bibnamefont
  {Lozovik}}\ and\ \bibinfo {author} {\bibfnamefont {V.~I.}\ \bibnamefont
  {Yudson}},\ }\bibfield  {title} {\bibinfo {title} {Feasibility of
  superfluidity of paired spatially separated electrons and holes; a new
  superconductivity mechanism},\ }\bibfield  {journal} {\bibinfo  {journal}
  {JETP Lett. (USSR) (Engl. Transl.); (United States)}\ }\href
  {https://www.osti.gov/biblio/7285279} {}\BibitemShut {NoStop}%
\bibitem [{\citenamefont {Pogrebinskii}(1977)}]{pogrebinskii1977mutual}%
  \BibitemOpen
  \bibfield  {author} {\bibinfo {author} {\bibfnamefont {M.}~\bibnamefont
  {Pogrebinskii}},\ }\bibfield  {title} {\bibinfo {title} {Mutual drag of
  carriers in a semiconductor-insulator-semiconductor system},\ }\href@noop {}
  {\bibfield  {journal} {\bibinfo  {journal} {Soviet Physics-Semiconductors}\
  }\textbf {\bibinfo {volume} {11}},\ \bibinfo {pages} {372} (\bibinfo {year}
  {1977})}\BibitemShut {NoStop}%
\bibitem [{\citenamefont {Blatt}\ \emph {et~al.}(1962)\citenamefont {Blatt},
  \citenamefont {B{\"o}er},\ and\ \citenamefont {Brandt}}]{blatt1962bose}%
  \BibitemOpen
  \bibfield  {author} {\bibinfo {author} {\bibfnamefont {J.~M.}\ \bibnamefont
  {Blatt}}, \bibinfo {author} {\bibfnamefont {K.}~\bibnamefont {B{\"o}er}},\
  and\ \bibinfo {author} {\bibfnamefont {W.}~\bibnamefont {Brandt}},\
  }\bibfield  {title} {\bibinfo {title} {Bose-einstein condensation of
  excitons},\ }\href@noop {} {\bibfield  {journal} {\bibinfo  {journal}
  {Physical Review}\ }\textbf {\bibinfo {volume} {126}},\ \bibinfo {pages}
  {1691} (\bibinfo {year} {1962})}\BibitemShut {NoStop}%
\bibitem [{\citenamefont {Kellogg}\ \emph {et~al.}(2004)\citenamefont
  {Kellogg}, \citenamefont {Eisenstein}, \citenamefont {Pfeiffer},\ and\
  \citenamefont {West}}]{kellogg2004vanishing}%
  \BibitemOpen
  \bibfield  {author} {\bibinfo {author} {\bibfnamefont {M.}~\bibnamefont
  {Kellogg}}, \bibinfo {author} {\bibfnamefont {J.}~\bibnamefont {Eisenstein}},
  \bibinfo {author} {\bibfnamefont {L.}~\bibnamefont {Pfeiffer}},\ and\
  \bibinfo {author} {\bibfnamefont {K.}~\bibnamefont {West}},\ }\bibfield
  {title} {\bibinfo {title} {Vanishing hall resistance at high magnetic field
  in a double-layer two-dimensional electron system},\ }\href@noop {}
  {\bibfield  {journal} {\bibinfo  {journal} {Physical review letters}\
  }\textbf {\bibinfo {volume} {93}},\ \bibinfo {pages} {036801} (\bibinfo
  {year} {2004})}\BibitemShut {NoStop}%
\bibitem [{\citenamefont {Su}\ and\ \citenamefont
  {MacDonald}(2008)}]{su2008make}%
  \BibitemOpen
  \bibfield  {author} {\bibinfo {author} {\bibfnamefont {J.-J.}\ \bibnamefont
  {Su}}\ and\ \bibinfo {author} {\bibfnamefont {A.}~\bibnamefont {MacDonald}},\
  }\bibfield  {title} {\bibinfo {title} {How to make a bilayer exciton
  condensate flow},\ }\href@noop {} {\bibfield  {journal} {\bibinfo  {journal}
  {Nature Physics}\ }\textbf {\bibinfo {volume} {4}},\ \bibinfo {pages} {799}
  (\bibinfo {year} {2008})}\BibitemShut {NoStop}%
\bibitem [{\citenamefont {Ju}\ \emph {et~al.}(2017)\citenamefont {Ju},
  \citenamefont {Wang}, \citenamefont {Cao}, \citenamefont {Taniguchi},
  \citenamefont {Watanabe}, \citenamefont {Louie}, \citenamefont {Rana},
  \citenamefont {Park}, \citenamefont {Hone}, \citenamefont {Wang},\ and\
  \citenamefont {McEuen}}]{Ju2017}%
  \BibitemOpen
  \bibfield  {author} {\bibinfo {author} {\bibfnamefont {L.}~\bibnamefont
  {Ju}}, \bibinfo {author} {\bibfnamefont {L.}~\bibnamefont {Wang}}, \bibinfo
  {author} {\bibfnamefont {T.}~\bibnamefont {Cao}}, \bibinfo {author}
  {\bibfnamefont {T.}~\bibnamefont {Taniguchi}}, \bibinfo {author}
  {\bibfnamefont {K.}~\bibnamefont {Watanabe}}, \bibinfo {author}
  {\bibfnamefont {S.~G.}\ \bibnamefont {Louie}}, \bibinfo {author}
  {\bibfnamefont {F.}~\bibnamefont {Rana}}, \bibinfo {author} {\bibfnamefont
  {J.}~\bibnamefont {Park}}, \bibinfo {author} {\bibfnamefont {J.}~\bibnamefont
  {Hone}}, \bibinfo {author} {\bibfnamefont {F.}~\bibnamefont {Wang}},\ and\
  \bibinfo {author} {\bibfnamefont {P.~L.}\ \bibnamefont {McEuen}},\ }\bibfield
   {title} {\bibinfo {title} {Tunable excitons in bilayer graphene},\ }\href
  {https://doi.org/10.1126/science.aam9175} {\bibfield  {journal} {\bibinfo
  {journal} {Science}\ }\textbf {\bibinfo {volume} {358}},\ \bibinfo {pages}
  {907} (\bibinfo {year} {2017})},\ \Eprint
  {https://arxiv.org/abs/https://www.science.org/doi/pdf/10.1126/science.aam9175}
  {https://www.science.org/doi/pdf/10.1126/science.aam9175} \BibitemShut
  {NoStop}%
\bibitem [{\citenamefont {Park}\ and\ \citenamefont {Louie}(2010)}]{Park2010}%
  \BibitemOpen
  \bibfield  {author} {\bibinfo {author} {\bibfnamefont {C.-H.}\ \bibnamefont
  {Park}}\ and\ \bibinfo {author} {\bibfnamefont {S.~G.}\ \bibnamefont
  {Louie}},\ }\bibfield  {title} {\bibinfo {title} {Tunable excitons in biased
  bilayer graphene},\ }\href {https://doi.org/10.1021/nl902932k} {\bibfield
  {journal} {\bibinfo  {journal} {Nano Letters}\ }\textbf {\bibinfo {volume}
  {10}},\ \bibinfo {pages} {426} (\bibinfo {year} {2010})}\BibitemShut
  {NoStop}%
\bibitem [{\citenamefont {Li}\ and\ \citenamefont
  {Appelbaum}(2019)}]{Appelbaum2019}%
  \BibitemOpen
  \bibfield  {author} {\bibinfo {author} {\bibfnamefont {P.}~\bibnamefont
  {Li}}\ and\ \bibinfo {author} {\bibfnamefont {I.}~\bibnamefont {Appelbaum}},\
  }\bibfield  {title} {\bibinfo {title} {Excitons without effective mass:
  Biased bilayer graphene},\ }\href
  {https://doi.org/10.1103/PhysRevB.99.035429} {\bibfield  {journal} {\bibinfo
  {journal} {Phys. Rev. B}\ }\textbf {\bibinfo {volume} {99}},\ \bibinfo
  {pages} {035429} (\bibinfo {year} {2019})}\BibitemShut {NoStop}%
\bibitem [{\citenamefont {Sauer}\ and\ \citenamefont
  {Pedersen}(2022)}]{Sauer2022}%
  \BibitemOpen
  \bibfield  {author} {\bibinfo {author} {\bibfnamefont {M.~O.}\ \bibnamefont
  {Sauer}}\ and\ \bibinfo {author} {\bibfnamefont {T.~G.}\ \bibnamefont
  {Pedersen}},\ }\bibfield  {title} {\bibinfo {title} {Exciton absorption, band
  structure, and optical emission in biased bilayer graphene},\ }\href
  {https://doi.org/10.1103/PhysRevB.105.115416} {\bibfield  {journal} {\bibinfo
   {journal} {Phys. Rev. B}\ }\textbf {\bibinfo {volume} {105}},\ \bibinfo
  {pages} {115416} (\bibinfo {year} {2022})}\BibitemShut {NoStop}%
\bibitem [{\citenamefont {Henriques}\ \emph {et~al.}(2022)\citenamefont
  {Henriques}, \citenamefont {Epstein},\ and\ \citenamefont
  {Peres}}]{Henriques2022}%
  \BibitemOpen
  \bibfield  {author} {\bibinfo {author} {\bibfnamefont {J.~C.~G.}\
  \bibnamefont {Henriques}}, \bibinfo {author} {\bibfnamefont {I.}~\bibnamefont
  {Epstein}},\ and\ \bibinfo {author} {\bibfnamefont {N.~M.~R.}\ \bibnamefont
  {Peres}},\ }\bibfield  {title} {\bibinfo {title} {Absorption and optical
  selection rules of tunable excitons in biased bilayer graphene},\ }\href
  {https://doi.org/10.1103/PhysRevB.105.045411} {\bibfield  {journal} {\bibinfo
   {journal} {Phys. Rev. B}\ }\textbf {\bibinfo {volume} {105}},\ \bibinfo
  {pages} {045411} (\bibinfo {year} {2022})}\BibitemShut {NoStop}%
\bibitem [{\citenamefont {Berestetskii}\ \emph {et~al.}(1982)\citenamefont
  {Berestetskii}, \citenamefont {Lifshitz},\ and\ \citenamefont
  {Pitaevskii}}]{berestetskii1982quantum}%
  \BibitemOpen
  \bibfield  {author} {\bibinfo {author} {\bibfnamefont {V.}~\bibnamefont
  {Berestetskii}}, \bibinfo {author} {\bibfnamefont {E.}~\bibnamefont
  {Lifshitz}},\ and\ \bibinfo {author} {\bibfnamefont {L.}~\bibnamefont
  {Pitaevskii}},\ }\href {https://books.google.com.au/books?id=URL5NKX8vbAC}
  {\emph {\bibinfo {title} {Quantum Electrodynamics: Volume 4}}},\ Course of
  theoretical physics\ (\bibinfo  {publisher} {Elsevier Science},\ \bibinfo
  {year} {1982})\BibitemShut {NoStop}%
\bibitem [{\citenamefont {Glazov}\ and\ \citenamefont
  {Chernikov}(2018)}]{Glazov2018}%
  \BibitemOpen
  \bibfield  {author} {\bibinfo {author} {\bibfnamefont {M.~M.}\ \bibnamefont
  {Glazov}}\ and\ \bibinfo {author} {\bibfnamefont {A.}~\bibnamefont
  {Chernikov}},\ }\bibfield  {title} {\bibinfo {title} {Breakdown of the static
  approximation for free carrier screening of excitons in monolayer
  semiconductors},\ }\href@noop {} {\bibfield  {journal} {\bibinfo  {journal}
  {Physica Status Solidi B}\ }\textbf {\bibinfo {volume} {255}},\ \bibinfo
  {pages} {180021} (\bibinfo {year} {2018})}\BibitemShut {NoStop}%
\bibitem [{\citenamefont {Zhang}\ \emph {et~al.}(2009)\citenamefont {Zhang},
  \citenamefont {Tang}, \citenamefont {Girit}, \citenamefont {Hao},
  \citenamefont {Martin}, \citenamefont {Zettl}, \citenamefont {Crommie},
  \citenamefont {Shen},\ and\ \citenamefont {Wang}}]{Zhang2009}%
  \BibitemOpen
  \bibfield  {author} {\bibinfo {author} {\bibfnamefont {Y.}~\bibnamefont
  {Zhang}}, \bibinfo {author} {\bibfnamefont {T.-T.}\ \bibnamefont {Tang}},
  \bibinfo {author} {\bibfnamefont {C.}~\bibnamefont {Girit}}, \bibinfo
  {author} {\bibfnamefont {Z.}~\bibnamefont {Hao}}, \bibinfo {author}
  {\bibfnamefont {M.~C.}\ \bibnamefont {Martin}}, \bibinfo {author}
  {\bibfnamefont {A.}~\bibnamefont {Zettl}}, \bibinfo {author} {\bibfnamefont
  {M.~F.}\ \bibnamefont {Crommie}}, \bibinfo {author} {\bibfnamefont {Y.~R.}\
  \bibnamefont {Shen}},\ and\ \bibinfo {author} {\bibfnamefont
  {F.}~\bibnamefont {Wang}},\ }\bibfield  {title} {\bibinfo {title} {Direct
  observation of a widely tunable bandgap in bilayer graphene},\ }\href
  {https://doi.org/10.1038/nature08105} {\bibfield  {journal} {\bibinfo
  {journal} {Nature}\ }\textbf {\bibinfo {volume} {459}},\ \bibinfo {pages}
  {820} (\bibinfo {year} {2009})}\BibitemShut {NoStop}%
\bibitem [{\citenamefont {Weitz}\ \emph {et~al.}(2010)\citenamefont {Weitz},
  \citenamefont {Allen}, \citenamefont {Feldman}, \citenamefont {Martin},\ and\
  \citenamefont {Yacoby}}]{yacoby2010}%
  \BibitemOpen
  \bibfield  {author} {\bibinfo {author} {\bibfnamefont {R.~T.}\ \bibnamefont
  {Weitz}}, \bibinfo {author} {\bibfnamefont {M.~T.}\ \bibnamefont {Allen}},
  \bibinfo {author} {\bibfnamefont {B.~E.}\ \bibnamefont {Feldman}}, \bibinfo
  {author} {\bibfnamefont {J.}~\bibnamefont {Martin}},\ and\ \bibinfo {author}
  {\bibfnamefont {A.}~\bibnamefont {Yacoby}},\ }\bibfield  {title} {\bibinfo
  {title} {Broken-symmetry states in doubly gated suspended bilayer graphene},\
  }\href {https://doi.org/10.1126/science.1194988} {\bibfield  {journal}
  {\bibinfo  {journal} {Science}\ }\textbf {\bibinfo {volume} {330}},\ \bibinfo
  {pages} {812} (\bibinfo {year} {2010})},\ \Eprint
  {https://arxiv.org/abs/https://www.science.org/doi/pdf/10.1126/science.1194988}
  {https://www.science.org/doi/pdf/10.1126/science.1194988} \BibitemShut
  {NoStop}%
\bibitem [{\citenamefont {Freitag}\ \emph {et~al.}(2012)\citenamefont
  {Freitag}, \citenamefont {Trbovic}, \citenamefont {Weiss},\ and\
  \citenamefont {Sch\"onenberger}}]{Freitag2012}%
  \BibitemOpen
  \bibfield  {author} {\bibinfo {author} {\bibfnamefont {F.}~\bibnamefont
  {Freitag}}, \bibinfo {author} {\bibfnamefont {J.}~\bibnamefont {Trbovic}},
  \bibinfo {author} {\bibfnamefont {M.}~\bibnamefont {Weiss}},\ and\ \bibinfo
  {author} {\bibfnamefont {C.}~\bibnamefont {Sch\"onenberger}},\ }\bibfield
  {title} {\bibinfo {title} {Spontaneously gapped ground state in suspended
  bilayer graphene},\ }\href {https://doi.org/10.1103/PhysRevLett.108.076602}
  {\bibfield  {journal} {\bibinfo  {journal} {Phys. Rev. Lett.}\ }\textbf
  {\bibinfo {volume} {108}},\ \bibinfo {pages} {076602} (\bibinfo {year}
  {2012})}\BibitemShut {NoStop}%
\bibitem [{\citenamefont {McCann}\ and\ \citenamefont
  {Fal'ko}(2006)}]{Falko2006}%
  \BibitemOpen
  \bibfield  {author} {\bibinfo {author} {\bibfnamefont {E.}~\bibnamefont
  {McCann}}\ and\ \bibinfo {author} {\bibfnamefont {V.~I.}\ \bibnamefont
  {Fal'ko}},\ }\bibfield  {title} {\bibinfo {title} {Landau-level degeneracy
  and quantum hall effect in a graphite bilayer},\ }\href
  {https://doi.org/10.1103/PhysRevLett.96.086805} {\bibfield  {journal}
  {\bibinfo  {journal} {Phys. Rev. Lett.}\ }\textbf {\bibinfo {volume} {96}},\
  \bibinfo {pages} {086805} (\bibinfo {year} {2006})}\BibitemShut {NoStop}%
\bibitem [{\citenamefont {McCann}\ and\ \citenamefont
  {Koshino}(2013)}]{McCann2013}%
  \BibitemOpen
  \bibfield  {author} {\bibinfo {author} {\bibfnamefont {E.}~\bibnamefont
  {McCann}}\ and\ \bibinfo {author} {\bibfnamefont {M.}~\bibnamefont
  {Koshino}},\ }\bibfield  {title} {\bibinfo {title} {The electronic properties
  of bilayer graphene},\ }\href {https://doi.org/10.1088/0034-4885/76/5/056503}
  {\bibfield  {journal} {\bibinfo  {journal} {Reports on Progress in Physics}\
  }\textbf {\bibinfo {volume} {76}},\ \bibinfo {pages} {056503} (\bibinfo
  {year} {2013})}\BibitemShut {NoStop}%
\bibitem [{\citenamefont {Kuzmenko}\ \emph {et~al.}(2009)\citenamefont
  {Kuzmenko}, \citenamefont {Crassee}, \citenamefont {van~der Marel},
  \citenamefont {Blake},\ and\ \citenamefont {Novoselov}}]{Kuzmenko2009}%
  \BibitemOpen
  \bibfield  {author} {\bibinfo {author} {\bibfnamefont {A.~B.}\ \bibnamefont
  {Kuzmenko}}, \bibinfo {author} {\bibfnamefont {I.}~\bibnamefont {Crassee}},
  \bibinfo {author} {\bibfnamefont {D.}~\bibnamefont {van~der Marel}}, \bibinfo
  {author} {\bibfnamefont {P.}~\bibnamefont {Blake}},\ and\ \bibinfo {author}
  {\bibfnamefont {K.~S.}\ \bibnamefont {Novoselov}},\ }\bibfield  {title}
  {\bibinfo {title} {Determination of the gate-tunable band gap and
  tight-binding parameters in bilayer graphene using infrared spectroscopy},\
  }\href {https://doi.org/10.1103/PhysRevB.80.165406} {\bibfield  {journal}
  {\bibinfo  {journal} {Phys. Rev. B}\ }\textbf {\bibinfo {volume} {80}},\
  \bibinfo {pages} {165406} (\bibinfo {year} {2009})}\BibitemShut {NoStop}%
\bibitem [{\citenamefont {Hwang}\ and\ \citenamefont
  {Das~Sarma}(2008)}]{Hwang2008}%
  \BibitemOpen
  \bibfield  {author} {\bibinfo {author} {\bibfnamefont {E.~H.}\ \bibnamefont
  {Hwang}}\ and\ \bibinfo {author} {\bibfnamefont {S.}~\bibnamefont
  {Das~Sarma}},\ }\bibfield  {title} {\bibinfo {title} {Screening, kohn
  anomaly, friedel oscillation, and rkky interaction in bilayer graphene},\
  }\href {https://doi.org/10.1103/PhysRevLett.101.156802} {\bibfield  {journal}
  {\bibinfo  {journal} {Phys. Rev. Lett.}\ }\textbf {\bibinfo {volume} {101}},\
  \bibinfo {pages} {156802} (\bibinfo {year} {2008})}\BibitemShut {NoStop}%
\end{thebibliography}%

\end{document}